\documentclass[twocolumn,showkeys,aps,prb,showpacs]{revtex4-1}
\usepackage{graphicx}
\usepackage[CJKbookmarks,dvipdfm,colorlinks,linkcolor=blue,citecolor=blue]{hyperref}

\begin{document}

\title{Coexistence of intrinsic piezoelectricity, ferromagnetism and nontrivial band topology  in Li-decorated Janus  monolayer $\mathrm{Fe_2SSe}$ with high  Curie temperature}

\author{San-Dong Guo$^{1}$, Wen-Qi Mu$^{1}$,  Meng-Yuan Yin$^{1}$, Yu-Chen Li$^{1}$   and  Wencai Ren$^{2,3}$}
\affiliation{$^1$School of Electronic Engineering, Xi'an University of Posts and Telecommunications, Xi'an 710121, China}
\affiliation{$^2$Shenyang National Laboratory for Materials Science, Institute of Metal Research, Chinese Academy of Sciences, Shenyang 110016, China}
\affiliation{$^3$School of Materials Science and Engineering, University of Science and Technology of China, Shenyang 110016, China}
\begin{abstract}
Recently, the quantum anomalous Hall (QAH) insulators are predicted  in Lithium-decorated
iron-based superconductor monolayer materials (LiFeX (X=S, Se and Te)) with very high  Curie temperature (\textcolor[rgb]{0.00,0.00,1.00}{PRL 125, 086401 (2020)}), which combines the topological and ferromagnetic (FM) orders.  It is interesting and useful to achieve coexistence of intrinsic piezoelectricity, ferromagnetism and nontrivial band topology in single two-dimensional (2D) material, namely 2D piezoelectric quantum anomalous hall insulator (PQAHI). In this work, 2D Janus monolayer  $\mathrm{Li_2Fe_2SSe}$  is  predict to be a room-temperature PQAHI, which possesses  dynamic, mechanical and thermal  stabilities. It is
predicted to be a half Dirac semimetal without spin-orbit coupling (SOC). It is found that the inclusion of SOC
opens up a large nontrivial gap, which means  the
nontrivial bulk topology (QAH insulator),  confirmed by the calculation of Berry curvature and the presence of two  chiral edge states (Chern number C=2). Calculated results show that  monolayer $\mathrm{Li_2Fe_2SSe}$  possesses robust QAH states against biaxial strain and electronic correlations. Compared to LiFeX, the glide mirror $G_z$ of $\mathrm{Li_2Fe_2SSe}$ disappears, which will induce  only out-of-plane piezoelectric response.  The calculated  out-of-plane $d_{31}$ of  monolayer $\mathrm{Li_2Fe_2SSe}$  is -0.238 pm/V comparable with ones of  other 2D known materials.
Moreover, very high  Curie temperature (about 1000 K) is predicted by  using Monte Carlo (MC)
simulations, which means that the QAH effect can be achieved at room temperature in Janus monolayer $\mathrm{Li_2Fe_2SSe}$.
 Similar to  monolayer $\mathrm{Li_2Fe_2SSe}$, the PQAHI can also  be realized  in the Janus monolayer $\mathrm{Li_2Fe_2SeTe}$.
 Our works open a new avenue in searching for
PQAHI with high temperature and high Chern numbers, which  provide a potential platform for multi-functional spintronic applications.

\end{abstract}

\pacs{71.20.-b, 77.65.-j, 72.15.Jf, 78.67.-n ~~~~~~~~~~~~~~~~~~~~~~~~~~~~~~~~~~~Email:sandongyuwang@163.com}

\maketitle

\section{Introduction}
The quantum Hall (QH) effect can be achieved in a 2D electron gas by  a
strong perpendicular external magnetic field\cite{qh1}, while  the occurrence of the QAH  effect aries from SOC and time-reversal  (TR) symmetry  broken in the presence of magnetic order\cite{qh2}.
The QAH effect is generally confirmed by a nonzero
Chern number in accordance with the number of edge states , and  only one spin species are allowed to flow unidirectionally,
resulting in a quantized Hall conductance\cite{zhj1,zhj2,zhj3}. The discovery of TR invariant
topological insulators  promotes the experimental realization of QAH effect\cite{qh3,qh4}.
Following the theoretical work\cite{qh5},  the QAH effect is achieved experimentally in thin films of Cr doped
$\mathrm{(Bi, Sb)_2Te}$ below 30 mK, with quantized Hall conductance being
observed\cite{qh6}. The QAH effect achieves  the coexistence of magnetism and topological
electronic band structure  in a single compound.  It's a natural idea to combine more properties in a material, like the coexistence of QAH effect and piezoelectricity.

The  piezoelectricity of 2D materials has
been widely investigated in recent years \cite{z}.
The  piezoelectricity of 2D materials has been observed experimentally\cite{q5,q6,q8,q8-1}, and the density functional theory (DFT) calculations have also predicted
the piezoelectric properties of many 2D materials \cite{q7-0,q7-1,q7-2,q7-4,q7-7,q7-8,q7-10,q9-0,q9-1,q9}, which lack centrosymmetry.
Recently, some  2D multifunctional piezoelectric  materials have been predicted by the first-principle calculations. In the 2D vanadium dichalcogenides,  $\mathrm{VSi_2P_4}$, $\mathrm{CrBr_{1.5}I_{1.5}}$ and $\mathrm{InCrTe_3}$\cite{qt1,q15,q15-1,q15-2},
the  piezoelectric ferromagnetism (PFM) has been predicted, which combines piezoelectricity and ferromagnetism.
The combination of piezoelectricity with   topological
insulating phase, namely piezoelectric quantum spin Hall insulator (PQSHI), has also been realized in monolayer  InXO (X=Se and Te)\cite{gsd1} and Janus monolayer $\mathrm{SrAlGaSe_4}$\cite{gsd2}.
These discoveries provide possibility  for using piezoelectric effect to control the quantum or spin transport process, which  may induce novel device applications or scientific breakthroughs.

\begin{figure*}
  \includegraphics[width=16.0cm]{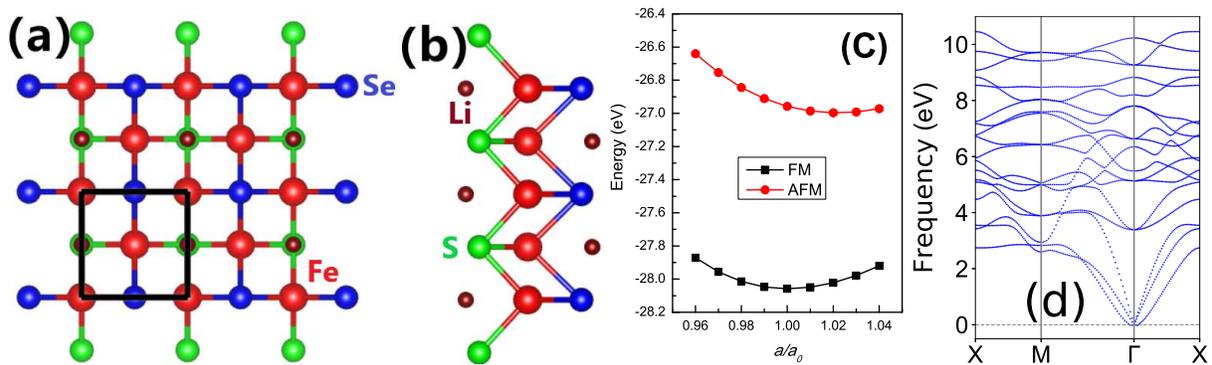}
  \caption{(Color online)The (a) top view and (b) side view of  crystal structure of Janus monolayer $\mathrm{Li_2Fe_2SSe}$.  The  black frame represents the  primitive cell. (c):The FM and AFM energy of  Janus monolayer  $\mathrm{Li_2Fe_2SSe}$ as a function of  $a/a_0$. (d):The phonon band dispersions  of Janus monolayer $\mathrm{Li_2Fe_2SSe}$  with FM  order.}\label{t0}
\end{figure*}

 In fact, in our recent work, the PQAHI has also been achieved  in Janus monolayer  $\mathrm{Fe_2IX}$ (X=Cl and Br)\cite{gsd22}, based on  QAH insulator  $\mathrm{Fe_2I_2}$ \cite{fe}.  Recently, the lithium decoration of layered iron-based
superconductor materials FeX (X=S, Se and Te), namely LiFeX, are  predicted as room-temperature QAH insulators with  high Chern number (C=2)\cite{pr}, and the guidances on experimental realization have also been discussed. The LiFeX (X=S, Se and Te) have similar crystal structure with  $\mathrm{Fe_2I_2}$, and  no piezoelectricity can be observed due to inversion symmetry.
A natural idea is  to achieve PQAHI from LiFeX (X=S, Se and Te) monolayer by removing inversion symmetry. As an example, the monolayer FeSe has particular sandwiched structure (Se-Fe-Se), and Janus structure can be built by replacing the top Se atomic layer  with S atoms ($\mathrm{Fe_2SSe}$), and then apply the gating techniques to inject a large amount of Li ions into monolayer $\mathrm{Fe_2SSe}$ ($\mathrm{Li_2Fe_2SSe}$). The related experimental  techniques have been widely used\cite{p1,p1-11,p1-12,p1-13,p1-14}.

In this work,   it is found  that, by first-principles calculations, Janus monolayer  $\mathrm{Li_2Fe_2SSe}$  is  a 2D ferromagnetic semiconductor with out-of-plane magnetization, which can achieve  the QAH effect at quite high temperature.
Janus monolayer  $\mathrm{Li_2Fe_2SSe}$ is proved to be dynamically, mechanically  and thermally stable.
 According to the results of Berry curvature and Chiral
edge states, a high Chern number C=2 is obtained  for $\mathrm{Li_2Fe_2SSe}$.
It is found that QAH effect of $\mathrm{Li_2Fe_2SSe}$  is  robust  against biaxial strain and electronic correlations.
  A very  high Curie temperature of
about 1000 K is  estimated by MC simulations using Heisenberg  model.
 A particular symmetry leads to only  out-of-plane piezoelectric response, and the predicted  out-of-plane $d_{31}$  is -0.238  pm/V, which  is  comparable with ones of  other 2D known materials.
 These results  indicate that 2D Janus   $\mathrm{Li_2Fe_2SSe}$
 may be promising candidate  for realizing the
room-temperature PQAHI in experiments, which is very useful for developing 2D piezoelectric spin topological devices.

The rest of the paper is organized as follows. In the next
section, we shall give our computational details and methods.
 In  the next few sections,  we shall present crystal structure, structural stabilities,  topological properties, strain and  correlation effects on  topological properties,   Curie temperature and piezoelectric properties  of 2D Janus   $\mathrm{Li_2Fe_2SSe}$.  Finally, we shall give our discussion and conclusions.

\begin{figure}
  \includegraphics[width=8cm]{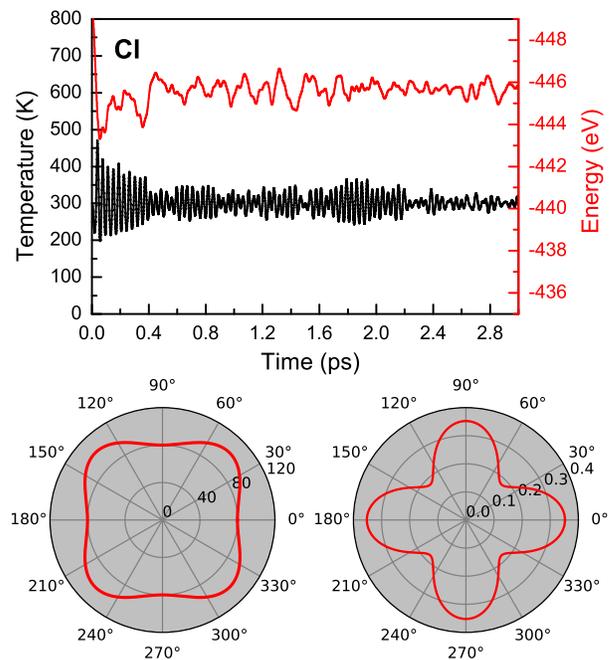}
\caption{(Color online)Top: the temperature and total energy fluctuations of  Janus monolayer  $\mathrm{Li_2Fe_2SSe}$  with FM  order at 300 K. Bottom: the angular dependence of the Young's modulus ($C_{2D}(\theta)$) and Poisson's
ratio ($\nu_{2D}(\theta)$) of Janus monolayer $\mathrm{Li_2Fe_2SSe}$.}\label{md}
\end{figure}

\section{Computational detail}
All the calculations on the elastic, electronic, topological and piezoelectric properties
are based on DFT\cite{1}, as implemented in the VASP code\cite{pv1,pv2,pv3}.
The projector-augmented-wave (PAW) potential and the
plane-wave basis with a kinetic energy cutoff of 500 eV are
employed for Janus monolayer $\mathrm{Li_2Fe_2SSe}$.
 We use popular generalized gradient approximation (GGA) of Perdew, Burke and  Ernzerhof\cite{pbe} as the exchange-correlation functional, and the SOC is included by a second variational procedure on a fully self-consistent basis.
 The total energy  convergence criterion is set  for $10^{-8}$ eV with the Gaussian smearing method. The force convergence criterion is set  for less than 0.0001 $\mathrm{eV.{\AA}^{-1}}$ for  optimizing the lattice constants and atomic coordinates  by
the conjugate gradient (CG) scheme.
To avoid artificial interactions caused by the periodic boundary condition,  the vacuum layer is set to more than 15 $\mathrm{{\AA}}$.
 The Fe-3$d$ orbitals  generally have important correlation effects, and  the
DFT+$U$ method\cite{u} is employed for the treatment of the strongly
correlated $3d$ electrons with $U_{eff}$ = 2.5 eV\cite{fe,fe1}.

With FM ground state, the interatomic force constants (IFCs) with the 5$\times$5$\times$1 supercell are calculated
  through the direct supercell method.  Based on  harmonic IFCs, the
phonon dispersions are attained by using Phonopy code\cite{pv5}.
An effective tight-binding Hamiltonian constructed
from the maximally localized Wannier function (MLWF) is
used to calculate  the surface states and Berry curvature with the iterative Green
function method, as implemented in the package WannierTools\cite{w1,w2}.
  The Curie temperature is
determined by MC simulations, as implemented by Mcsolver code\cite{mc}.

The elastic stiffness tensor  $C_{ij}$ and  piezoelectric stress tensor $e_{ij}$  are carried out by using strain-stress relationship (SSR) with GGA and  density functional perturbation theory (DFPT) method\cite{pv6} using GGA+SOC.
 The Brillouin zone (BZ) integration is sampled by using a
18$\times$18$\times$1 Monkhorst-Pack grid for the self-consistent calculations   and elastic coefficients $C_{ij}$.    A very dense  mesh of 26$\times$26$\times$1 k-points in the BZ is adopted to attain the accurate $e_{ij}$.
The 2D elastic coefficients $C^{2D}_{ij}$
 and   piezoelectric stress coefficients $e^{2D}_{ij}$
have been renormalized by   $C^{2D}_{ij}$=$Lz$$C^{3D}_{ij}$ and $e^{2D}_{ij}$=$Lz$$e^{3D}_{ij}$, where the $Lz$ is  the length of unit cell along z direction.

\section{Crystal Structure}
The FeSe monolayer has Se-Fe-Se trilayers with a tetragonal lattice
in the $P4/nmm$ space group, the unit cell of which  contains four atoms with two co-planar Fe atoms.
It is proved that the room-temperature  QAH insulator can be achieved by Li decoration of FeSe monolayer\cite{pr}. In fact, the $\mathrm{Fe_2I_2}$ monolayer, having   the same
crystal structure with FeSe, is predicted to be a QAH insulator\cite{fe}. The I element has one more valence electron than Se element, and the Li atom with an ultralow electronegativity easily loses one valence. So, the  Li decoration of FeSe monolayer can become a QAH insulator.
It is well known that Janus monolayer MoSSe (S-Mo-Se) can be constructed from $\mathrm{MoSe_2}$ with three atomic sublayers (Se-Mo-Se)\cite{p1}. In our previous work, the Janus monolayer  $\mathrm{Fe_2IX}$ (X=Cl and Br)  is predicted to be PQAHI\cite{gsd22}, which is built by  replacing one of two I  layers with X (X=Cl and Br) atoms in monolayer  $\mathrm{Fe_2I_2}$. It is a natural idea to construct Janus $\mathrm{Fe_2SSe}$  monolayer (S-Fe-Se)   by  replacing one of two Se  layers with S  atoms in monolayer  $\mathrm{FeSe}$, and then to achieve PQAHI by  Li decoration of Janus monolayer $\mathrm{Fe_2SSe}$ ($\mathrm{Li_2Fe_2SSe}$).
Compared with FeSe monolayer, the $\mathrm{Li_2Fe_2SSe}$ monolayer with  $P4mm$ space group (No.99) lacks centrosymmetry, giving rise to  piezoelectricity. The top and side views of schematic crystal structure of Janus monolayer $\mathrm{Li_2Fe_2SSe}$ are shown in \autoref{t0}.

 The ground state of  Janus monolayer $\mathrm{Li_2Fe_2SSe}$ can be determined by comparing the energy difference between antiferromagnetic (AFM) and FM states. The  magnetic ground state of $\mathrm{Li_2Fe_2SSe}$ monolayer  is FM, which can be seen from the (c) in \autoref{t0}.  The optimized lattice constants $a_0$ is 3.636 $\mathrm{{\AA}}$ with FM order, which is between ones of LiFeS (3.542  $\mathrm{{\AA}}$) and LiFeSe (3.655  $\mathrm{{\AA}}$)\cite{pr}. The thermal stability of magnetic
ordering can be described by the  magnetic anisotropy energy (MAE), stemming  from the
SOC effect.  The (100) and  (001) directions are used to obtain relative stabilities by using GGA+SOC.
 The energy difference of the magnetic moments constrained in the (100) and
(001) direction is  172 $\mu eV$/Fe,  indicating that the out-of-plane
(001) direction is the easy one for magnetization in monolayer $\mathrm{Li_2Fe_2SSe}$.
A built-in electric field can be produced by the the inequivalent bond lengths and  bond  angles (see \autoref{tab}) due to different atomic sizes and electronegativities of S and Se atoms. For LiFeSe monolayer,  the key space-group symmetry operations contain
space inversion $P$, $C_4$ rotation, $M_x$ and $M_y$ mirrors and
glide mirror $G_z= \{M_z|\frac{1}{2},\frac{1}{2},0\}$\cite{pr}. For $\mathrm{Li_2Fe_2SSe}$ monolayer, besides the missing $P$, the glide mirror $G_z$ is also removed.

\begin{figure}
  \includegraphics[width=8cm]{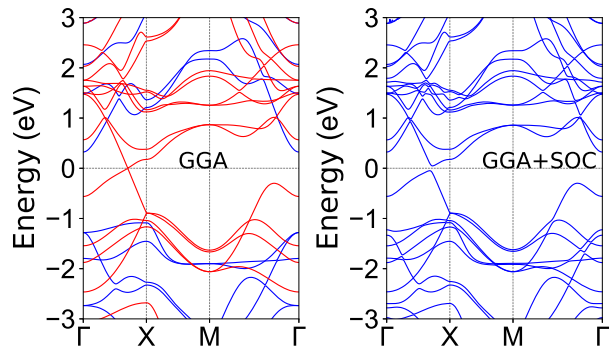}
  \caption{(Color online) The  energy band structures of  Janus monolayer  $\mathrm{Li_2Fe_2SSe}$ without and with SOC at the FM order. The blue (red) lines represent the  spin-up (spin-down) bands without SOC.  }\label{band}
\end{figure}
\begin{figure}
  \includegraphics[width=8cm]{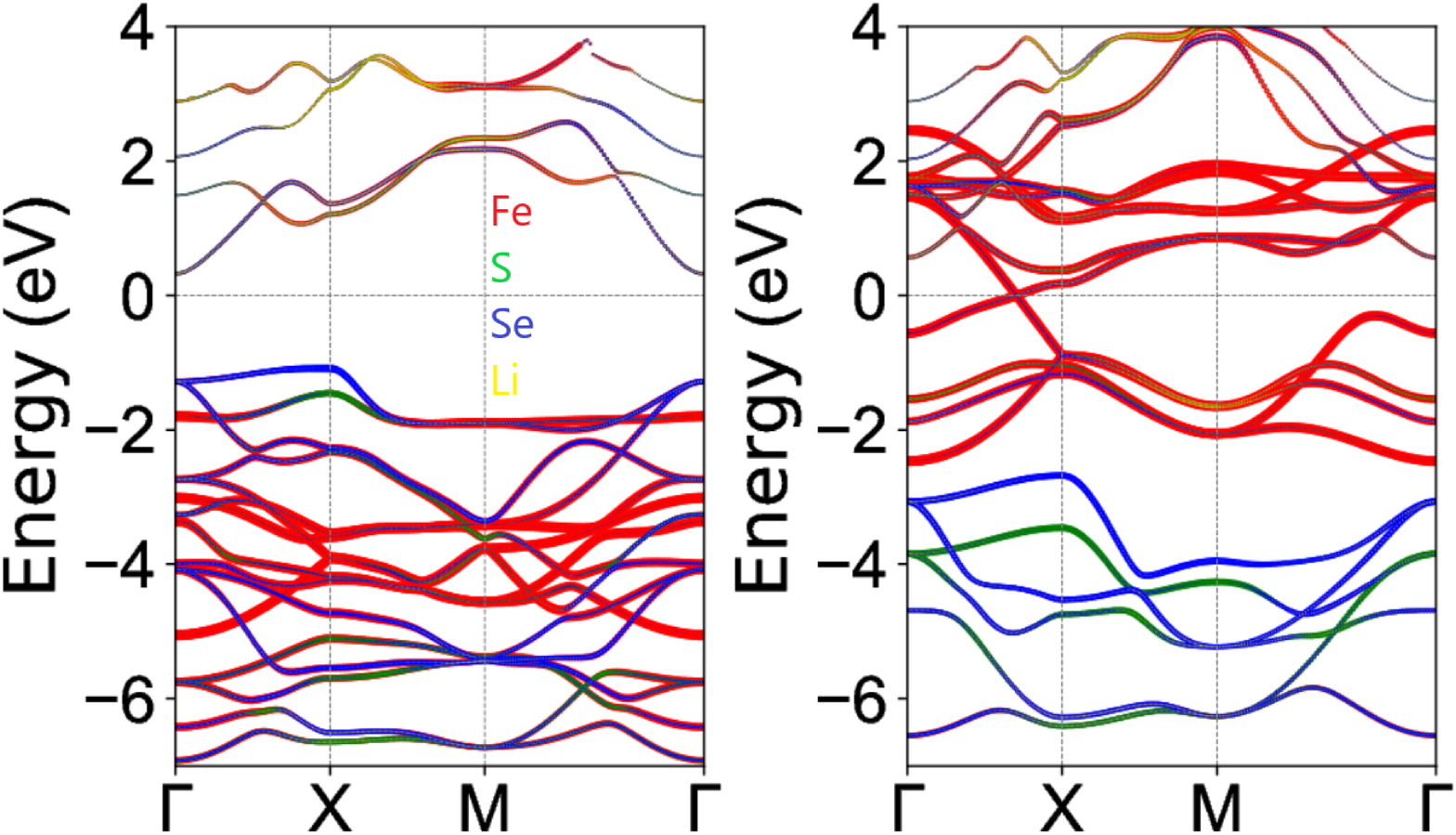}
  \caption{(Color online) Band structure of Janus monolayer  $\mathrm{Li_2Fe_2SSe}$ without the
SOC for spin up (left) and spin down (right).  The
contribution of Fe, S, Se and Li atoms to the Bloch
states are denoted by red, green, blue and yellow dots.  }\label{band-1}
\end{figure}
\begin{figure*}
  \includegraphics[width=15cm]{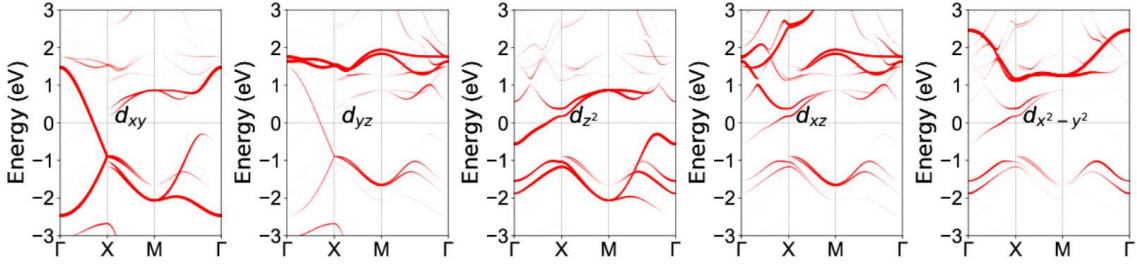}
  \caption{(Color online)At the absence of SOC,  the band structure of Janus monolayer  $\mathrm{Li_2Fe_2SSe}$  for spin down with the
contribution of $d_{xy}$, $d_{yz}$, $d_{z^2}$, $d_{xz}$ and $d_{x^2-y^2}$  orbitals to the Bloch
states.   }\label{band-2}
\end{figure*}

\begin{figure*}
  \includegraphics[width=12.0cm]{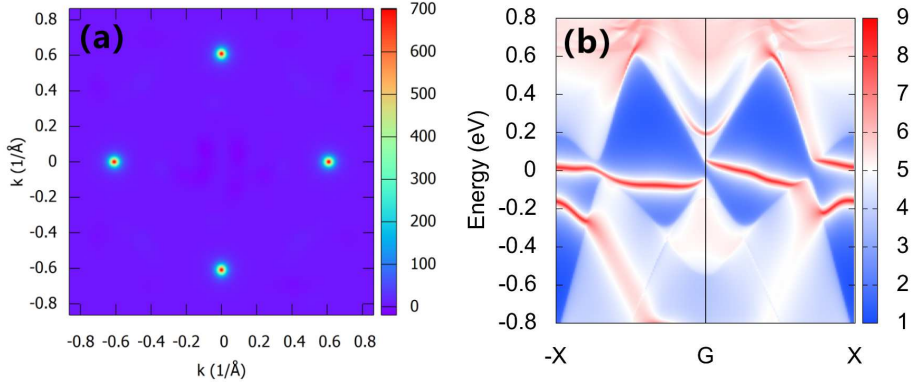}
  \caption{(Color online)(a): The distribution of Berry curvature of Janus monolayer  $\mathrm{Li_2Fe_2SSe}$  contributed
by occupied valence bands in the momentum space. (b): The topological
edge states of $\mathrm{Li_2Fe_2SSe}$  calculated  along the (100) direction.}\label{berry}
\end{figure*}

\begin{figure}
  \includegraphics[width=8cm]{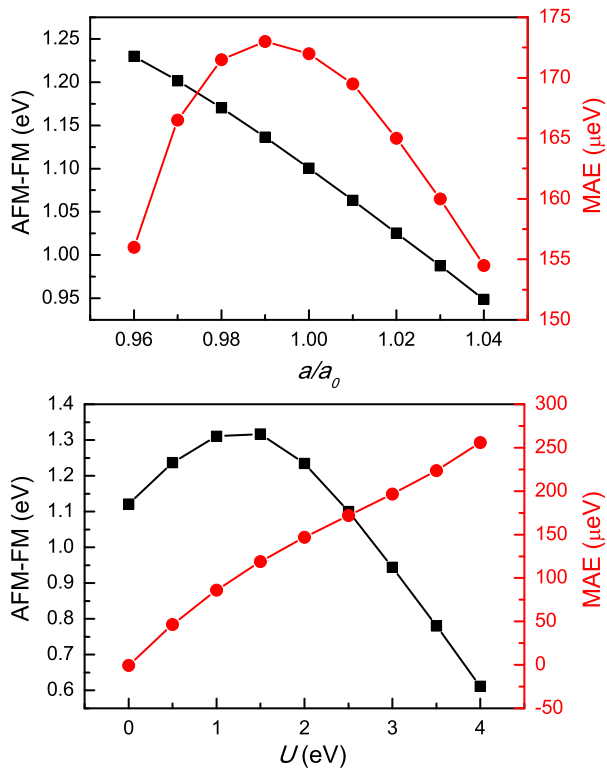}
\caption{(Color online)For Janus monolayer  $\mathrm{Li_2Fe_2SSe}$, the energy difference between  FM and AFM order and MAE  as a function of  $a/a_0$ (Top) and $U$ (Bottom). }\label{em}
\end{figure}

\begin{table}
\centering \caption{For Janus monolayer  $\mathrm{Li_2Fe_2SSe}$, the lattice constants $a_0$ ($\mathrm{{\AA}}$);  Fe-Se ($d_1$), Fe-S ($d_2$),  Li-Se ($d_3$), Li-S ($d_4$) bond lengths ($\mathrm{{\AA}}$); Se-Fe-Se ($\theta_1$) and S-Fe-S ($\theta_2$) angles;  the thickness layer height $t$ ($\mathrm{{\AA}}$);  the elastic constants $C_{ij}$ in $\mathrm{Nm^{-1}}$. }\label{tab}
  \begin{tabular*}{0.48\textwidth}{@{\extracolsep{\fill}}cccccc}
  \hline\hline
 $a_0$ &  $d_1$&$d_2$& $d_3$& $d_4$& $\theta_1$\\\hline
   3.636&2.587&2.467&2.599&2.612&89.26\\\hline\hline
  $\theta_2$& $t$&$C_{11}$& $C_{12}$& $C_{66}$&\\\hline
   94.94&4.347&91.26&32.15&43.00&\\\hline\hline
\end{tabular*}
\end{table}

\section{Structural Stability}
 The phonon
calculation of $\mathrm{Li_2Fe_2SSe}$ (see (d) in \autoref{t0}) reveals that there are no imaginary
frequency modes,  confirming its
dynamic stability, which means that $\mathrm{Li_2Fe_2SSe}$ monolayer can exist as free-standing
2D crystal.
A total of 18 branches due to 6
atoms per unitcell can be observed, including 15 optical and 3 acoustical phonon
branches.  The out-of-plane acoustic (ZA)  branch  (out-of-plane vibrations) deviates from linearity, in accord  with  the conclusion that  the  ZA phonon branch should have quadratic dispersion for  the  unstrained monolayer\cite{r1,r2}. The thermal stability of $\mathrm{Li_2Fe_2SSe}$ monolayer is further assessed by performing ab initio molecular dynamics (AIMD) simulations  using NVT ensemble with a  4$\times$4$\times$1  supercell at 300 K. For  $\mathrm{Li_2Fe_2SSe}$ monolayer,
the temperature and total energy fluctuations as a function of the simulation time are shown in \autoref{md}.
It is found that  the total energy and temperature fluctuate smoothly with small
amplitudes after the preheating process, which means  a favorable thermal stability for $\mathrm{Li_2Fe_2SSe}$ monolayer at room tempearure.

\begin{figure*}
    \includegraphics[width=15cm]{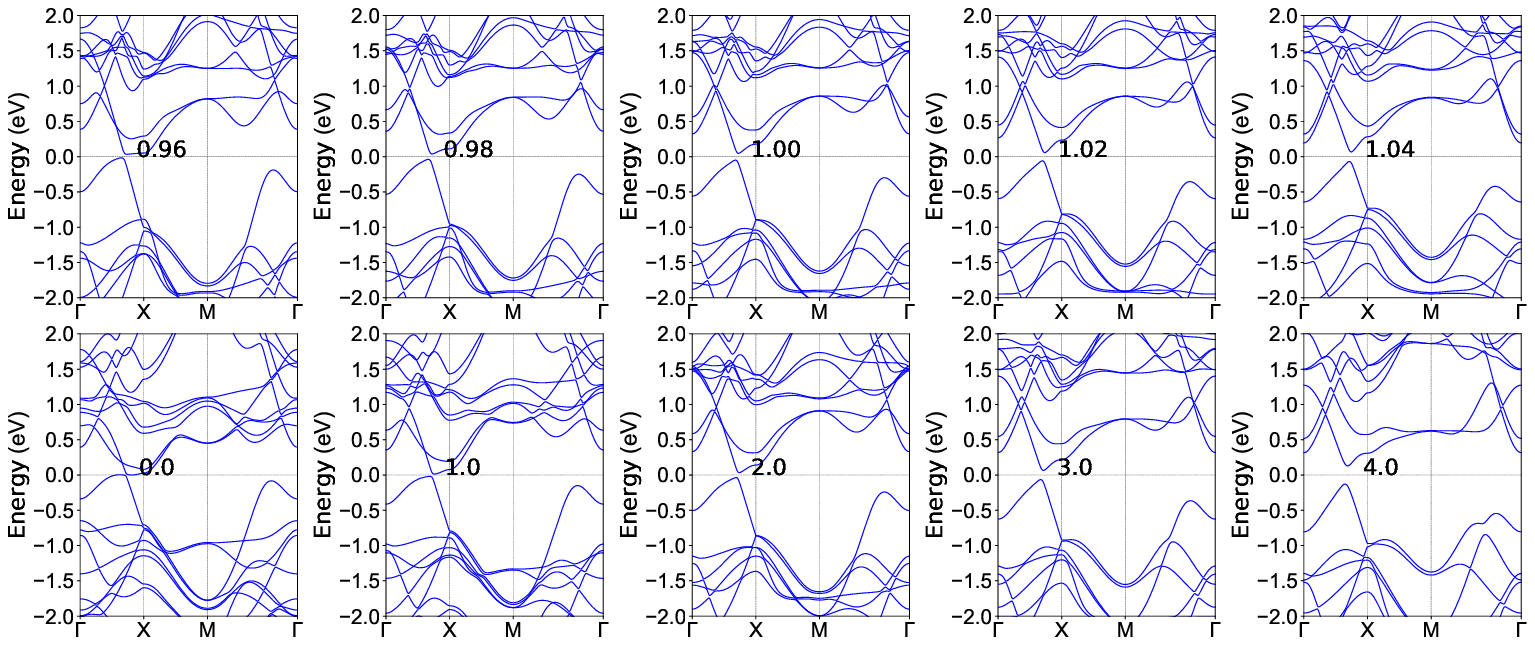}
  \caption{(Color online) The energy band structures  of Janus monolayer  $\mathrm{Li_2Fe_2SSe}$  using GGA+SOC with five  different $a/a_0$ (Top) and $U$ (Bottom).}\label{band-s}
\end{figure*}
\begin{figure}
   \includegraphics[width=8.0cm]{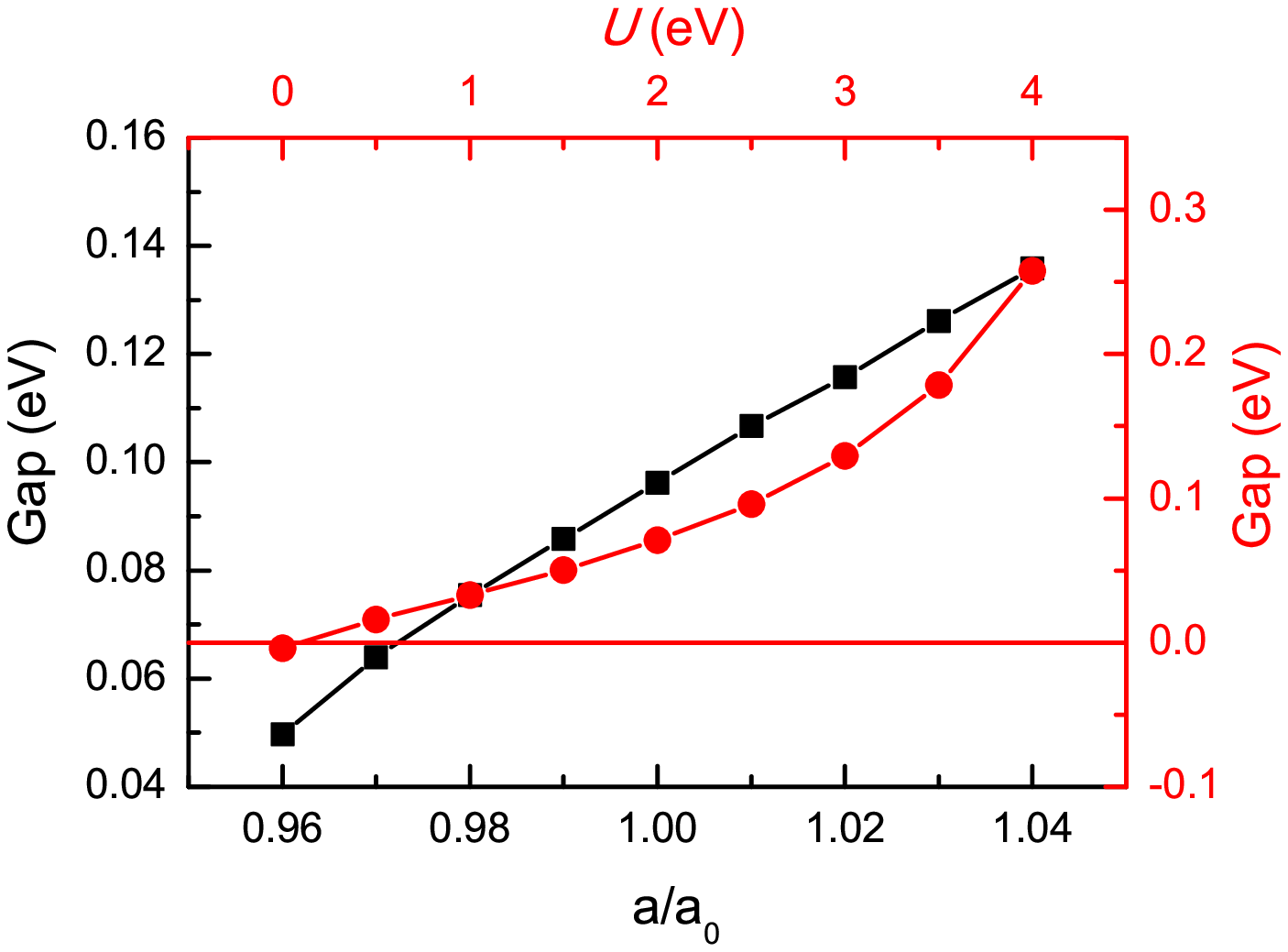}
  \caption{(Color online) For Janus monolayer  $\mathrm{Li_2Fe_2SSe}$, the gap as a function of $a/a_0$ and $U$  using GGA+SOC.}\label{gap-s}
\end{figure}

The mechanical stability of  monolayer  $\mathrm{Li_2Fe_2SSe}$ can be checked by Born  criteria of mechanical stability:
 \begin{equation}\label{pe1-4}
  C_{11}>0,~~ C_{66}>0,~~C_{11}-C_{12}>0
\end{equation}
where the $C_{11}$, $C_{12}$ and $C_{66}$ are three independent elastic
constants. By using Voigt notation, the elastic tensor with $4mm$ point-group symmetry  can be reduced into:
\begin{equation}\label{pe1-4}
   C=\left(
    \begin{array}{ccc}
      C_{11} & C_{12} & 0 \\
     C_{12} & C_{11} &0 \\
      0 & 0 & C_{66} \\
    \end{array}
  \right)
\end{equation}
The calculated $C_{11}$, $C_{12}$ and  $C_{66}$ are 91.26 $\mathrm{Nm^{-1}}$, 32.15 $\mathrm{Nm^{-1}}$ and 43.00 $\mathrm{Nm^{-1}}$, which
satisfy the above Born  criteria of mechanical stability,  indicating  mechanical stability of  monolayer  $\mathrm{Li_2Fe_2SSe}$.

Due to $C_4$ rotation symmetry, the  mechanical properties of monolayer $\mathrm{Li_2Fe_2SSe}$ have  $C_4$ symmetry, and
the direction-dependent  in-plane Young's moduli $C_{2D}(\theta)$ and
Poisson's ratios $\nu_{2D}(\theta)$ can be attained by \cite{ela,ela1}:
 \begin{equation}\label{pe1-4-1}
  C_{2D}(\theta)=\frac{C_{11}C_{22}-C_{12}^2}{C_{11}m^4+C_{22}n^4+(B-2C_{12})m^2n^2}
\end{equation}
 \begin{equation}\label{pe1-4-2}
  \nu_{2D}(\theta)=\frac{(C_{11}+C_{22}-B)m^2n^2-C_{12}(m^4+n^4)}{C_{11}m^4+C_{22}n^4+(B-2C_{12})m^2n^2}
\end{equation}
in which  $m=sin(\theta)$, $n=cos(\theta)$ and $B=(C_{11}C_{22}-C_{12}^2)/C_{66}$. The $\theta$ is the angle of the direction with
 the x direction  as $0^{\circ}$.
The  $C_{2D}(\theta)$ and
 $\nu_{2D}(\theta)$  as a function of the angle $\theta$ are plotted in \autoref{md}. It is clearly seen that they show  $C_4$ symmetry, and
 we only consider the $0^{\circ}$-$90^{\circ}$ angle range.
 The softest
direction  is along the (100)
direction, while
the hardest direction is
 along the (110) direction, and the corresponding Young's
moduli is  79.94 $\mathrm{Nm^{-1}}$  and 101.36 $\mathrm{Nm^{-1}}$.
The maximum value of Young's moduli is
less than that of many 2D materials\cite{q7-0,q7-4,gra}, which means that the  monolayer  $\mathrm{Li_2Fe_2SSe}$ has extraordinary
flexibility.
For Poisson's ratios,  the minima   is  along the
(110) direction (0.179), while the maxima is along
the (100) direction (0.352).

\section{Topological properties}
The energy band structures of monolayer  $\mathrm{Li_2Fe_2SSe}$ with GGA is plotted in \autoref{band}, along with atom projected band structure in \autoref{band-1}. Without SOC,   monolayer  $\mathrm{Li_2Fe_2SSe}$ shows  a  2D half Dirac semimetal state with a large-gap insulator for spin up (1.41 eV) and a gapless Dirac semimetal for spin down. Due to  $C_4$ symmetry, the four Dirac cones in the BZ can be observed for spin down along the mirror symmetry invariant lines $\Gamma$-X and $\Gamma$-Y.
The important
observation is that the states around the Fermi level for  spin down are
dominated by the Fe-$d$ orbitals, which means that the Fe-$d$ orbitals are partially occupied.
However, the spin-up channel
is fully occupied.  These lead to the high-spin  state for Fe atom,
giving a spin magnetic moment of 3 $\mu_B$. To understand
the  composition of  Dirac cone, we project the
states to five Fe-$d$ orbitals for spin down, which are plotted in \autoref{band-2}.
 Calculated results show that the states around Dirac cone are dominated by the $d_{xy}$ and  $d_{z^2}$ orbital.  We then investigate the electronic band structures with SOC, and the energy bands are shown in \autoref{band}.
When including  SOC,  the Dirac point  splits apart
with a noticeable energy gap opened, and the corresponding gap is 96.2 meV for  $\mathrm{Li_2Fe_2SSe}$, which suggests nontrivial
topology.

When the TR symmetry of a material  breaks with a
finite magnetic ordering, the topologically nontrivial properties can be verified by a non-zero Chern number
in the valence bands.
The Chern number of monolayer  $\mathrm{Li_2Fe_2SSe}$ can be attained by integrating the Berry curvature ($\Omega_z(k)$) of the
occupied bands:
 \begin{equation}\label{pe1-4}
  C=\frac{1}{2\pi}\int_{BZ}d^2k \Omega_z(k)
\end{equation}
\begin{equation}\label{pe1-4}
  \Omega_z(k)=\nabla_k\times i\langle\mu_{n,k}|\nabla_k\mu_{n,k}\rangle
\end{equation}
in which the  $\mu_{n,k}$ is the lattice periodic part of the Bloch wave functions.
The distributions of
Berry curvature of  monolayer  $\mathrm{Li_2Fe_2SSe}$ are shown in \autoref{berry}.
As can be observed, the nonzero Berry curvature is mainly distributed around four Dirac cones, and they  have the same sign because of $C_4$ symmetry. Two
Berry curvature peaks  contribute to the nonzero Chern number 1, and  the total Chern number
from the 4 Berry curvature peaks equals 2. In other words,  a quantized
Berry phase of $\pi$ can be attained for each gapped Dirac cone, and the total Berry phase of 4$\pi$ is attained (four Dirac cones), giving rise to a Chern number C=2.

Based on the bulk-edge correspondence,  the non-zero
Chern number can be further confirmed by the number of nontrivial
chiral edge states  inside the bulk gap of a semiinfinite system.
The local density of states vs momentum and energy  at the edge can be
obtained from the imaginary part of the surface Green's
function:
 \begin{equation}\label{pe1-4}
  A(k,\omega)=-\frac{1}{\pi}\lim_{\eta\rightarrow 0^+} \mathrm{ImTr} G_s(k, \omega+i\eta)
\end{equation}
The topological
edge states  along the (100) direction  are shown in \autoref{berry}.
It is clearly seen that  the bulk states are
connected by two chiral edge states, which
indicates that  Chern number equals to 2.
These results show that monolayer  $\mathrm{Li_2Fe_2SSe}$ is a QAH insulator.

\begin{figure*}
  \includegraphics[width=12.0cm]{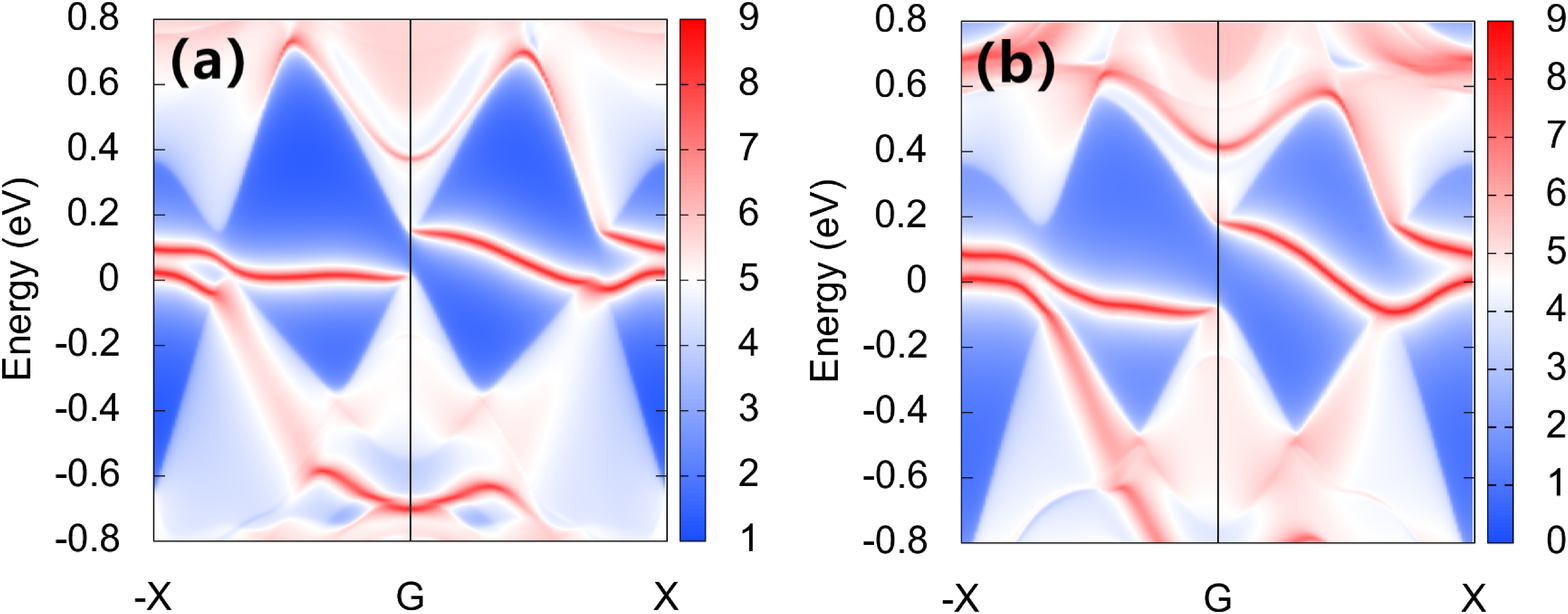}
  \caption{(Color online)Topological
edge states of Janus monolayer  $\mathrm{Li_2Fe_2SSe}$ calculated  along the (100) direction with $a/a_0$=1.04 (a) and $U$=4 eV (b).}\label{ss-s}
\end{figure*}
\begin{figure}
   \includegraphics[width=8cm]{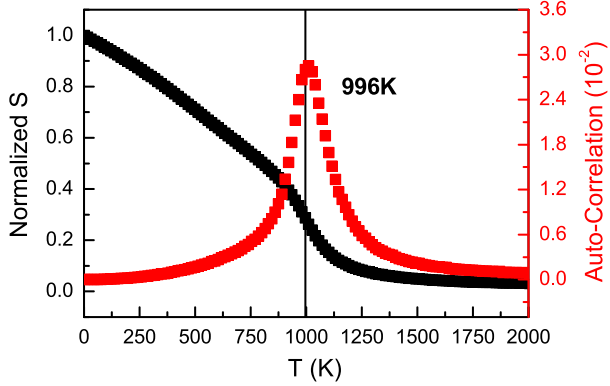}
    \caption{(Color online)The normalized magnetic moment (S) and auto-correlation of Janus monolayer  $\mathrm{Li_2Fe_2SSe}$ as a function of temperature.  }\label{ed}
\end{figure}
\begin{figure}
   \includegraphics[width=7.0cm]{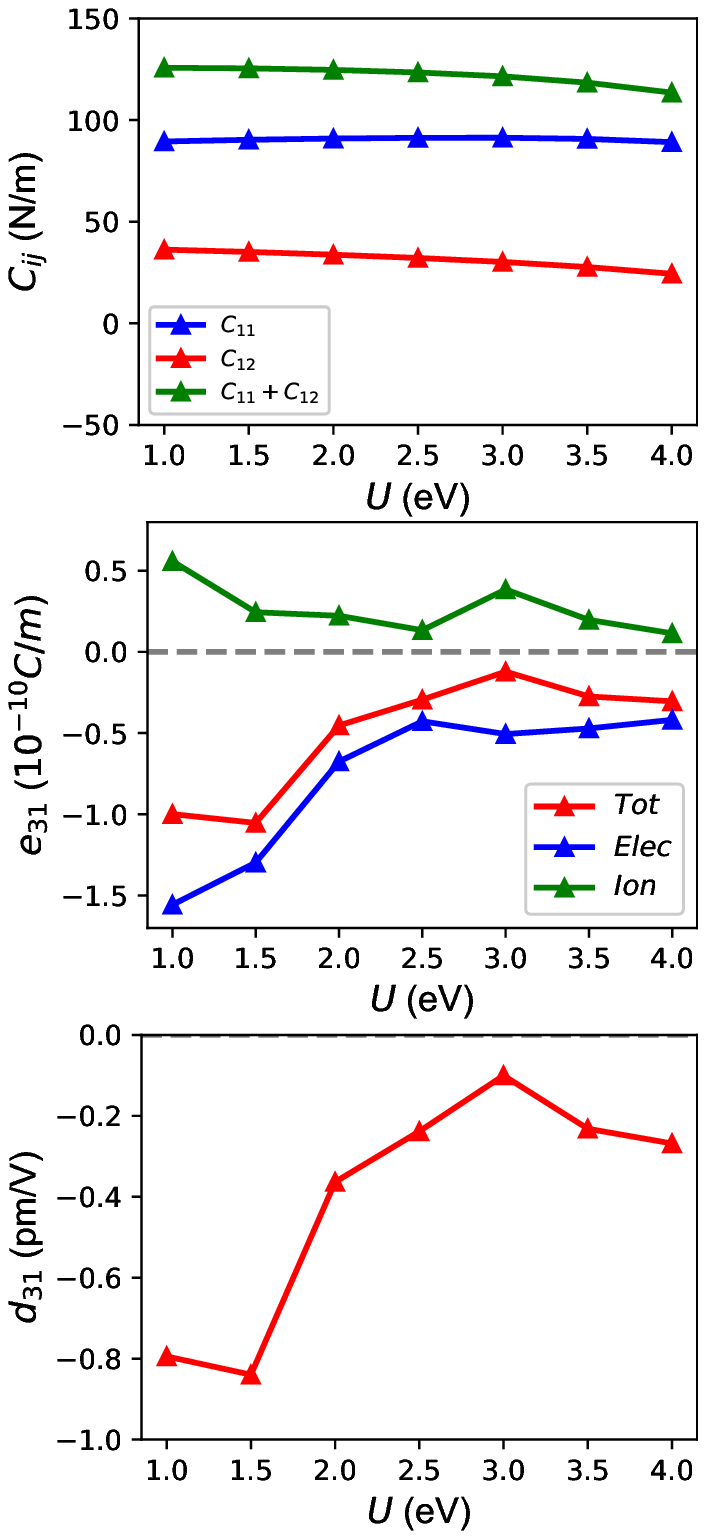}
  \caption{(Color online) For Janus monolayer  $\mathrm{Li_2Fe_2SSe}$, the elastic constants  $C_{ij}$, the piezoelectric stress coefficients  ($e_{31}$) and the piezoelectric strain coefficients  ($d_{31}$) as a function of $U$.}\label{p-u}
\end{figure}

\section{Strain and  correlation effects on  topological properties}
The strain generally can tune the SOC-induced bulk gap, MAE and magnetic order.
It is important to investigate robustness of QAH properties of monolayer  $\mathrm{Li_2Fe_2SSe}$
against biaxial  strain.  Here, we use  $a/a_0$ (0.96-1.04)  to describe the biaxial strain, where
$a$ ($a_0$) is the strained (equilibrium) lattice constants.
The energy difference between  FM and AFM orders and MAE  as a function of  $a/a_0$ are plotted in \autoref{em}.
It is found that  all strained monolayers are FM ground state in considered strain range, and the energy difference between  FM and AFM order decreases with $a/a_0$ from 0.96 to 1.04. However, the MAE  shows nonmonotonicity, which  firstly increases, and then decreases. In considered strain range,  the out-of-plane (001) direction is always the easy one.
The energy band structures  of some representative strained monolayer  $\mathrm{Li_2Fe_2SSe}$ using GGA+SOC   are plotted in \autoref{band-s}, and the gap as a function of strain is shown in \autoref{gap-s}.
 It is found that the gap  increases almost linearly with increasing strain in considered strain range, and  the gap changes from 49.7 meV to 135.8 meV. The topological
edge states of strained monolayer $\mathrm{Li_2Fe_2SSe}$ are  calculated, and they all are QAH insulators with Chern number C=2.
We  show  topological
edge states  at representative 1.04 strain in \autoref{ss-s}, which shows clearly   two chiral topologically
protected gapless edge states.  These results show that the QAH topological properties of monolayer $\mathrm{Li_2Fe_2SSe}$ are robust  against strain.

To check the Coulomb interaction $U$  effects on
QAH properties, we calculate the energy difference between  FM and AFM orders, MAE,  energy  bands and topological
edge states  with different values of $U$ (0-4 eV). The energy difference between  FM and AFM order and MAE  as a function of $U$ are plotted in \autoref{em}. It is found that the ground state always is FM order with different $U$. It is found that the MAE increases with increasing $U$. When the $U$ value is larger than 0 eV, the out-of-plane (001) direction is always the easy one in considered $U$ range. For $U$=0 eV, the MAE only is -0.5 $\mu eV$/Fe, and  the spin orientation in the energy band calculations is chosen in the out-of-plane direction. The energy band structures  of  monolayer  $\mathrm{Li_2Fe_2SSe}$ with some representative $U$ value using GGA+SOC   are plotted in \autoref{band-s}, and the gap as a function of $U$ is shown in \autoref{gap-s}. With increasing $U$, it is clearly seen that the gap  increases from -4.1 meV to 257.6 meV. For U=0 eV, the  monolayer $\mathrm{Li_2Fe_2SSe}$ becomes metal.
In considered $U$ range except $U$=0 eV, the topological
edge states of monolayer $\mathrm{Li_2Fe_2SSe}$  are  calculated, and they all are QAH insulators with Chern number C=2.
The  topological
edge states  at representative $U$=4 eV are shown in \autoref{ss-s}, and   two chiral topologically
protected gapless edge states are present in the bulk gap. These results  indicate
the robustness of nontrivial topology against the correlation
effect in the 3$d$ electrons of Fe atoms.

\section{Curie temperature}
One of the important properties of ferromagnets for  the practical application is the Curie temperatures ($T_C$).
Using MC
simulations based on the Heisenberg model, we have calculated
the Curie temperature of monolayer $\mathrm{Li_2Fe_2SSe}$. For simplicity, only the nearest neighbor
(NN) exchange interaction is considered, and the spin Heisenberg Hamiltonian is defined as:
  \begin{equation}\label{pe0-1-1}
H=-J\sum_{i,j}S_i\cdot S_j-A\sum_i(S_i^z)^2
 \end{equation}
where  $J$, $S$ and $A$ are the  exchange parameter,  the
spin vector of each atom and   MAE, respectively.
Based on the energy
difference between AFM and FM, the magnetic coupling parameter
is calculated as $J$=($E_{AFM}$-$E_{FM}$)/8 with normalized
$S$ ($|S|$ = 1). The  calculated $J$ value is 137.6 meV, and the A is  172 $\mu eV$/Fe.

A 50$\times$50  supercell
with periodic boundary conditions is employed, and  $10^7$ loops are adopted to perfotme the
MC simulation. We show the
normalized magnetic moment and auto-correlation of monolayer $\mathrm{Li_2Fe_2SSe}$  as a function of temperature  in \autoref{ed}.
It is found that  $T_C$  is as
high as 996 K for  monolayer $\mathrm{Li_2Fe_2SSe}$, which is smaller than ones of Li-decorated monolayer FeX (X= S, Se and  Te)\cite{pr}.
However, the $T_C$  of  monolayer $\mathrm{Li_2Fe_2SSe}$ is  significantly higher than that of previously
reported many  2D FM semiconductors, like $\mathrm{CrI_3}$ monolayer (about 45 K)\cite{m7-6}, CrOCl monolayer
(about 160 K)\cite{tc1}, Janus $\mathrm{Fe_2IX}$ (X=Cl and Br) monolayer (about 400 K)\cite{gsd22},  $\mathrm{Fe_2I_2}$ monolayer (about 400 K)\cite{fe}  and $\mathrm{Cr_2Ge_2Te_6}$ monolayer (about 20 K)\cite{tc2}.

\section{Piezoelectric properties}
The piezoelectric effects of a material can be described by  third-rank piezoelectric stress tensor  $e_{ijk}$ and strain tensor $d_{ijk}$.   The $e_{ijk}$ and $d_{ijk}$ can be expressed as:
 \begin{equation}\label{pe0}
      e_{ijk}=\frac{\partial P_i}{\partial \varepsilon_{jk}}=e_{ijk}^{elc}+e_{ijk}^{ion}
 \end{equation}
and
 \begin{equation}\label{pe0-1}
   d_{ijk}=\frac{\partial P_i}{\partial \sigma_{jk}}=d_{ijk}^{elc}+d_{ijk}^{ion}
 \end{equation}
 They can be related  by elastic tensor $C_{mnjk}$:
 \begin{equation}\label{pe9-1-1}
    e_{ijk}=\frac{\partial P_i}{\partial \varepsilon_{jk}}=\frac{\partial P_i}{\partial \sigma_{mn}}.\frac{\partial \sigma_{mn}}{\partial\varepsilon_{jk}}=d_{imn}C_{mnjk}
 \end{equation}
in which  $P_i$, $\varepsilon_{jk}$ and $\sigma_{jk}$ are polarization vector, strain and stress, respectively.
The $e_{ijk}^{elc}$/$d_{ijk}^{elc}$ is  clamped-ion piezoelectric coefficients (only  electronic contributions).  The  $e_{ijk}$/$d_{ijk}$ is relax-ion piezoelectric coefficients as a realistic result (the sum of ionic
and electronic contributions).

 The Li-decorated monolayer FeX (X= S, Se and  Te)   are centrosymmetric, which means that they have no piezoelectricity. However,
 the monolayer $\mathrm{Li_2Fe_2SSe}$  lacks  glide mirror $G_z$ symmetry, but has $M_x$ and $M_y$ mirrors symmetry, which gives only out-of-plane piezoelectricity, and the  in-plane piezoelectricity will disappear.
By using  Voigt notation,    the  piezoelectric stress   and strain tensors of monolayer $\mathrm{Li_2Fe_2SSe}$ can be expressed as:
 \begin{equation}\label{pe1-1}
 e=\left(
    \begin{array}{ccc}
     0 & 0 & 0 \\
     0 & 0 & 0 \\
      e_{31} & e_{31} & 0 \\
    \end{array}
  \right)
    \end{equation}

  \begin{equation}\label{pe1-2}
  d= \left(
    \begin{array}{ccc}
      0 & 0 & 0 \\
       0 & 0 & 0 \\
      d_{31} & d_{31} &0 \\
    \end{array}
  \right)
\end{equation}
 The $e_{31}$ can be directly calculated by VASP code, and the
$d_{31}$  can be  derived by \autoref{pe9-1-1}, \autoref{pe1-1} and \autoref{pe1-2}.
\begin{equation}\label{pe2}
    d_{31}=\frac{e_{31}}{C_{11}+C_{12}}
\end{equation}

Next, we use the primitive cell  to calculate the  $e_{31}$  of  Janus monolayer  $\mathrm{Li_2Fe_2SSe}$.
Calculated results show that the $e_{31}$ is -0.294$\times$$10^{-10}$ C/m with  the  ionic contribution   0.133$\times$$10^{-10}$ C/m  and electronic one -0.427 $\times$$10^{-10}$ C/m. It is found  that the electronic and
ionic parts  have  the opposite contributions, and  the electronic  contribution
dominates the $e_{31}$. Based on \autoref{pe2}, the  calculated $d_{31}$ is -0.238 pm/V.
The $d_{31}$ of Janus monolayer  $\mathrm{Li_2Fe_2SSe}$ is  higher  than or comparable with ones of many 2D  materials\cite{o1,o2,o3,o4}.
The Coulomb interaction $U$  effects on
piezoelectric properties of  monolayer  $\mathrm{Li_2Fe_2SSe}$ are also considered. The elastic constants  $C_{ij}$, the piezoelectric stress coefficients  ($e_{31}$) and the piezoelectric strain coefficients  ($d_{31}$) as a function of $U$ are plotted in \autoref{p-u}.
A complex $U$ dependence for $e_{31}$ is observed, which leads to complicated $U$ effects on $d_{31}$.
It is found that the smallest $d_{31}$ is -0.100 pm/V with $U$=3 eV.

\begin{figure}
   \includegraphics[width=8cm]{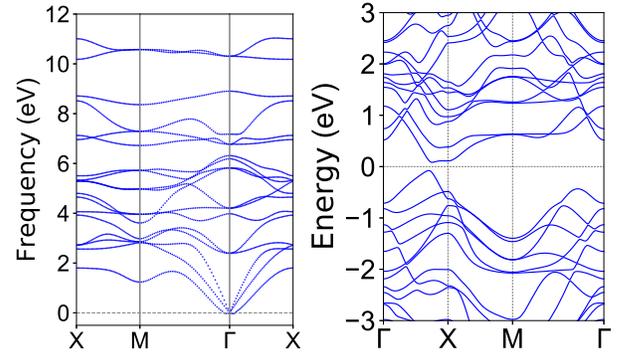}
  \caption{(Color online) For Janus monolayer  $\mathrm{Li_2Fe_2SeTe}$, the phonon spectra (Left) and the energy band structure with GGA+SOC (Right).}\label{p-sete}
\end{figure}

\section{Discussion and Conclusion}
\begin{figure}
   \includegraphics[width=7cm]{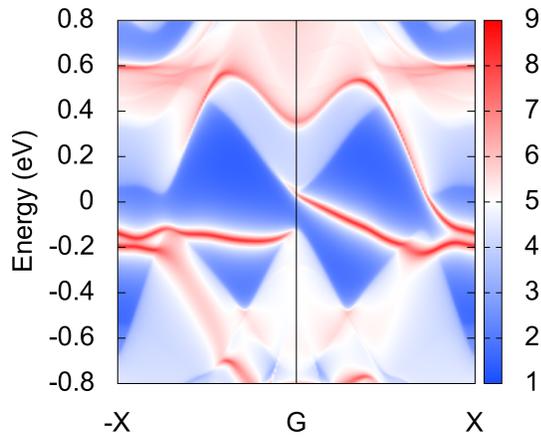}
  \caption{(Color online) For Janus monolayer  $\mathrm{Li_2Fe_2SeTe}$,  the topological edge states  calculated  along the (100) direction.}\label{p-sete-1}
\end{figure}

 In fact, Janus monolayer  $\mathrm{Li_2Fe_2SeTe}$ is also a QAH insulator.
 The energy difference between  AFM and FM orders is 0.875 eV per unitcell, and the out-of-plane (001) direction is  the easy one with MAE of 45 $\mu eV$/Fe. The calculated $C_{11}$, $C_{12}$ and  $C_{66}$ are 79.13 $\mathrm{Nm^{-1}}$, 19.06 $\mathrm{Nm^{-1}}$ and 31.36 $\mathrm{Nm^{-1}}$, which satisfy  Born  criteria of mechanical stability,  indicating  mechanical stability of  monolayer  $\mathrm{Li_2Fe_2SeTe}$.
 The  phonon spectra  of $\mathrm{Li_2Fe_2SeTe}$  is plotted in \autoref{p-sete}  with no imaginary
frequency modes,  confirming its dynamic stability. The AIMD simulations also confirm the thermal  stability of monolayer  $\mathrm{LiFe_2SeTSe}$ at room temperature.
The energy band structures and  topological
edge states of  monolayer $\mathrm{Li_2Fe_2SeTe}$ are shown in \autoref{p-sete} and \autoref{p-sete-1}, respectively. The energy band gap is 161 meV, and the bulk states are
connected by two chiral edge states,
indicating   Chern number C=2. The calculated $d_{31}$ is 0.1 pm/V with $e_{31}$ of 0.099$\times$$10^{-10}$ C/m.

In summary,  using  DFT+$U$ calculations, we have performed a
systematic investigation of the electronic, magnetic, topological and piezoelectric properties
in Janus monolayer  $\mathrm{Li_2Fe_2SSe}$, which is predicted as an  intriguing 2D  PQAHI.
It is proved  that monolayer  $\mathrm{Li_2Fe_2SSe}$ is mechanically, dynamically and thermally stable. Also, the nontrivial
properties  are confirmed by a nonzero
Chern number (C=2) and two gapless chiral edge states. Moreover, the monolayer  $\mathrm{Li_2Fe_2SSe}$ possesses out-of-plane magnetic anisotropy and
very high Curie temperature (about 1000 K).
The emergence of QAH effect in monolayer $\mathrm{Li_2Fe_2SSe}$ is robust against strain and  Hubbard $U$ electronic correlation.
The predicted $d_{31}$   is  comparable with ones of  other 2D known materials.
The predicted PQAHI in  monolayer  $\mathrm{Li_2Fe_2SSe}$ is expected to work safely above room
temperature, and provides a more promising
platform for realizing low-dissipation topotronics devices, and provides possibility to use the piezotronic effect to control QAH effects.

\begin{acknowledgments}
This work is supported by Natural Science Basis Research Plan in Shaanxi Province of China  (2021JM-456). We are grateful to the Advanced Analysis and Computation Center of China University of Mining and Technology (CUMT) for the award of CPU hours and WIEN2k/VASP software to accomplish this work.
\end{acknowledgments}


\begin{references}
\bibitem{qh1} K. V. Klitzing, G. Dorda and M. Pepper, Phys. Rev. Lett. \textbf{45}, 494 (1980).

\bibitem{qh2}F. D. M. Haldane, Phys. Rev. Lett. \textbf{61}, 2015 (1988).

\bibitem{zhj1}D. Q. Zhang, M. J. Shi, T. S. Zhu, D. Y. Xing, H. J. Zhang and J. Wang, Phys. Rev. Lett. \textbf{122}, 206401 (2019).

\bibitem{zhj2}H. J. Zhang, Y. Xu, J. Wang and S. C. Zhang, Phys. Rev. Lett. \textbf{112}, 216803 (2014).

\bibitem{zhj3}H. J. Zhang, J. Wang, G. Xu, Y. Xu and S. C. Zhang, Phys. Rev. Lett. \textbf{112}, 096804 (2014).

\bibitem{qh3}M. Z. Hasan and C. L. Kane, Rev. Mod. Phys. \textbf{82},
3045 (2010).

\bibitem{qh4}X. L. Qi and S. C. Zhang, Rev. Mod. Phys. \textbf{83}, 1057 (2011).



\bibitem{qh5}R. Yu, W. Zhang, H. J. Zhang, S. C. Zhang, X. Dai and
Z. Fang, Science \textbf{329}, 61 (2010).

\bibitem{qh6} C. Z. Chang, J. Zhang, X. Feng, J. Shen, Z. Zhang, M. Guo,
K. Li, Y. Ou, P. Wei, L. L. Wang, Z. Q. Ji, Y. Feng,
S. Ji, X. Chen, J. Jia, X. Dai, Z. Fang, S. C. Zhang, K. He,
Y. Wang, L. Lu, X. C. Ma and Q. K. Xue, Science \textbf{340},
167 (2013).

\bibitem{z}P. Lin, C. Pan and  Z. L. Wang, Materials Today Nano \textbf{4}, 17 (2018).



\bibitem{q5}M. Dai, Z. Wang, F. Wang, Y. Qiu, J. Zhang, C. Y. Xu, T. Zhai, W. Cao, Y. Fu,
D. Jia, Y. Zhou, and P. A. Hu, Nano Lett. \textbf{19}, 5416 (2019).

\bibitem{q6} W. Wu, L. Wang, Y. Li, F. Zhang, L. Lin, S. Niu, D. Chenet,
X. Zhang, Y. Hao, T. F. Heinz, J. Hone and Z. L. Wang,
Nature \textbf{514}, 470 (2014).

\bibitem{q8}A. Y. Lu, H. Zhu, J. Xiao, C. P. Chuu, Y. Han, M. H. Chiu,
C. C. Cheng, C. W. Yang, K. H. Wei, Y. Yang, Y. Wang,
D. Sokaras, D. Nordlund, P. Yang, D. A. Muller, M. Y. Chou,
X. Zhang and L. J. Li, Nat. Nanotechnol. \textbf{12}, 744 (2017).

\bibitem{q8-1}H. Zhu, Y. Wang, J. Xiao, M. Liu, S. Xiong, Z. J. Wong, Z. Ye,
Y. Ye, X. Yin and X. Zhang, Nat. Nanotechnol. \textbf{10},
151 (2015).


\bibitem{q7-0}L. Dong, J. Lou and V. B. Shenoy, ACS Nano, \textbf{11},
8242 (2017).

\bibitem{q7-1}R. X. Fei, We. B. Li, J. Li and L. Yang, Appl. Phys. Lett.  \textbf{107}, 173104 (2015).



\bibitem{q7-2}M. N. Blonsky, H. L. Zhuang, A. K. Singh and R.  G. Hennig,  ACS Nano, \textbf{9},
9885 (2015).

\bibitem{q7-4} S. D. Guo, Y. T. Zhu, W. Q. Mu and W. C. Ren,  EPL \textbf{132},  57002 (2020).


\bibitem{q7-7}W. B. Li  and J. Li, Nano Res.  \textbf{8}, 3796 (2015).


\bibitem{q7-8}Dimple, N. Jena, A. Rawat, R.  Ahammed,
M. K. Mohanta and A. D. Sarkar, J. Mater. Chem. A  \textbf{6},
24885 (2018).



\bibitem{q7-10}N. Jena, Dimple, S. D.  Behere  and A. D. Sarkar, J. Phys. Chem. C  \textbf{121}, 9181 (2017).

\bibitem{q9-0}M. T. Ong and E.J. Reed,  ACS Nano \textbf{6}, 1387 (2012).

\bibitem{q9-1}A. A. M. Noor, H. J. Kim  and Y. H. Shin, Phys. Chem. Chem. Phys. \textbf{16}, 6575 (2014).


\bibitem{q9}J. Tan, Y. H. Wang, Z. T. Wang, X. J. He, Y. L. Liu, B. Wanga, M. I. Katsnelson and  S. J.  Yuan, Nano Energy \textbf{65},  104058 (2019).

\bibitem{qt1}J. H. Yang,  A. P. Wang, S. Z. Zhang, J.  Liu, Z. C. Zhong and L. Chen, Phys. Chem. Chem. Phys.,
\textbf{21}, 132 (2019).

\bibitem{q15}S. D. Guo, W. Q. Mu, Y. T. Zhu and X. Q. Chen, Phys. Chem. Chem. Phys. \textbf{22}, 28359 (2020).

\bibitem{q15-1}S. D. Guo, X. S. Guo, X. X. Cai, W. Q. Mu and  W. C. Ren, 	arXiv:2103.15141 (2021).

\bibitem{q15-2}G. Song, D. S. Li, H. F. Zhou et al., Appl. Phys. Lett. \textbf{118}, 123102 (2021).


\bibitem{gsd1}S. D. Guo, W. Q. Mu, Y. T. Zhu, S. Q. Wang and  G. Z. Wang, 	J. Mater. Chem. C \textbf{9}, 5460 (2021).

\bibitem{gsd2}S. D. Guo, Y. T. Zhu, W. Q. Mu and X. Q. Chen, J. Mater. Chem. C   DOI: 10.1039/D1TC01165K (2021)..

\bibitem{gsd22}S. D. Guo, W. Q. Mu, X. B. Xiao and B. G. Liu, arXiv:2105.03004 (2021).


\bibitem{fe}Q. L. Sun,  Y. D. Ma and N. Kioussis, Mater. Horiz. \textbf{7}, 2071 (2020).

\bibitem{pr}Y. Li, J. H. Li, Y. Li, M. Ye, F. W. Zheng, Z. T. Zhang, J. H. Fu, W. H. Duan and Y. Xu, Phys. Rev. Lett. \textbf{125}, (2020).

\bibitem{p1}A. Y. Lu, H. Y. Zhu, J. Xiao et al., Nature Nanotechnology \textbf{12}, 744 (2017).



\bibitem{p1-11}Y. Deng, Y. Yu, Y. Song, J. Zhang, N. Z. Wang, Z. Sun, Y. Yi, Y. Z. Wu, S. Wu, J. Zhu, et al., Nature \textbf{563}, 94 (2018).

\bibitem{p1-12}L. Ma, B. Lei, N. Wang, K. Yang, D. Liu, F. Meng, C. Shang, Z. Sun, J. Cui, C. Zhu, et al., Sci. Bull. \textbf{64}, 653 (2019).

\bibitem{p1-13}B. Lei, L. Ma, N. Wang, C. Zhu, J. Cui, Z. Sun, D. Ma, H. Wang, M. Shi, J. Ying, et al., Phys. Rev. B \textbf{100}, 174519 (2019).

\bibitem{p1-14}E. Dagotto, Rev. Mod. Phys. \textbf{85}, 849 (2013).

\bibitem{1}P. Hohenberg and W. Kohn, Phys. Rev. \textbf{136},
B864 (1964); W. Kohn and L. J. Sham, Phys. Rev. \textbf{140},
A1133 (1965).

\bibitem{pv1} G. Kresse, J. Non-Cryst. Solids \textbf{193}, 222 (1995).

\bibitem{pv2} G. Kresse and J. Furthm$\ddot{u}$ller, Comput. Mater. Sci. 6, \textbf{15} (1996).

\bibitem{pv3} G. Kresse and D. Joubert, Phys. Rev. B \textbf{59}, 1758 (1999).
\bibitem{pbe}J. P. Perdew, K. Burke and M. Ernzerhof, Phys. Rev. Lett. \textbf{77}, 3865 (1996).

\bibitem{u}V. I. Anisimov, F. Aryasetiawan and A. I. Lichtenstein,
J. Phys. Condens. Mat. \textbf{9}, 767 (1997).

\bibitem{fe1}J. Heyd, G. E. Scuseria and M. Ernzerhof, J. Chem. Phys. \textbf{118}, 8207 (2003).




\bibitem{pv5}A. Togo, F. Oba, and I. Tanaka, Phys. Rev. B \textbf{78}, 134106
(2008).

\bibitem{w1} Q. Wu, S. Zhang, H. F. Song, M. Troyer and A. A. Soluyanov, Comput. Phys. Commun. \textbf{224}, 405
(2018).
\bibitem{w2}A. A. Mostofia, J. R. Yatesb, G. Pizzif, Y.-S. Lee, I. Souzad, D.
Vanderbilte and N. Marzarif,  Comput. Phys. Commun. \textbf{185}, 2309 (2014).


\bibitem{mc}L. Liu, X. Ren, J. H. Xie, B. Cheng, W. K. Liu, T. Y. An, H. W. Qin and  J. F. Hu, Appl. Surf. Sci.  \textbf{480},  300 (2019).
\bibitem{pv6}X. Wu, D. Vanderbilt and  D. R.  Hamann, Phys. Rev. B  \textbf{72}, 035105 (2005).

\bibitem{r1}E. Mariani and F. V. Oppen, Phys. Rev. Lett. \textbf{100}, 076801 (2008).


\bibitem{r2}J. Carrete , W. Li, L. Lindsay, D. A. Broido, L. J. Gallego and N. Mingo, Mater. Res. Lett. \textbf{4}, 204 (2016).


\bibitem{ela}E. Cadelano and L. Colombo, Phys. Rev. B  \textbf{85}, 245434 (2012).

\bibitem{ela1} E. Cadelano, P. L. Palla, S. Giordano, and L. Colombo, Phys.
Rev. B  \textbf{82}, 235414 (2010).

\bibitem{gra}C. Lee, X. Wei, J. W. Kysar, and J. Hone, Science \textbf{321}, 385
(2008).

\bibitem{m7-6}B. Huang, G. Clark, E. Navarro-Moratalla, D. R. Klein, R. Cheng,
K. L. Seyler, D. Zhong, E. Schmidgall, M. A. McGuire, D. H.
Cobden, W. Yao, D. Xiao, P. Jarillo-Herrero and X. Xu, Nature \textbf{546}, 270 (2017).


\bibitem{tc1}N. Miao, B. Xu, L. Zhu, J. Zhou and Z. Sun, J. Am. Chem. Soc. \textbf{140}, 2417 (2018).


\bibitem{tc2}C. Gong, L. Li, Z. Li, H. Ji, A. Stern, Y. Xia, T. Cao, W. Bao,
C. Wang, Y. Wang, Z. Q. Qiu, R. J. Cava, S. G. Louie, J. Xia
and X. Zhang, Nature  \textbf{546}, 265 (2017).



\bibitem{o1}A. A. M. Noor, H. J. Kim and Y. H. Shin,  Phys. Chem.
Chem. Phys. \textbf{16}, 6575 (2014).

\bibitem{o2} M. T. Ong and E. J. Reed, ACS Nano \textbf{6},  1387 (2012).

\bibitem{o3}Y. Guo, S. Zhou, Y. Z. Bai, and J. J. Zhao, Appl. Phys. Lett. \textbf{110}, 163102 (2017).


\bibitem{o4}L. Hu and X.R. Huang, RSC Adv. \textbf{7},  55034 (2017).

\end{references}
\end{document}